\documentclass[adp,a4paper,fleqn]{w-art}
\usepackage{times,cite,w-thm}
\theoremstyle{plain}

\theoremstyle{definition}

\usepackage[]{graphicx}
\usepackage{amsmath}
\usepackage{amsfonts}
\usepackage{amssymb}

\begin{document}

\DOIsuffix{theDOIsuffix}
\Volume{empty}
\Month{empty}
\Year{empty}
\pagespan{1}{}
\Receiveddate{XXXX}
\Reviseddate{XXXX}
\Accepteddate{XXXX}
\Dateposted{XXXX}
\keywords{space-time, gravitation, equations of gravitation.}
\subjclass[pacs]{
\qquad\parbox[t][2.2\baselineskip][t]{100mm}{%
  \raggedright
  04.20.Cv, 04.50.+h\vfill}}%

\title{Geodesic-invariant equations of gravitation}
\author{L. Verozub\footnote{Corresponding
     author: e-mail: {\sf l.verozub@googlemail.com}, 
    }
} 
\address[
]{Kharkov National University, 61077, Kharkov, Ukraine}

\begin{abstract}
 Einstein's equations
of gravitation are not invariant under geodesic
mappings, i. e.  under a certain class of mappings of the Christoffel symbols and the metric tensor  which leave
the geodesic equations  in a given coordinate system invariant. A theory in which geodesic mappings play the role of gauge transformations is considered.
\end{abstract}
\maketitle

\DOIsuffix{theDOIsuffix}

\Volume{} \Issue{} \Copyrightissue{} \Month{} \Year{}

\pagespan{3}{}

\Receiveddate{} \Reviseddate{} \Accepteddate{} \Dateposted{}

\section{Introduction}

  If we start with Einstein's beautiful hypothesis that test particles
in gravitational field move along geodesic lines of some Riemannian spaces
$V_{4}$, it is natural to expect that the differential equations for finding
the metric tensor $g_{\alpha\beta}(x)$ for a given distribution of matter also should
be invariant under any transformations at which the geodesic equations remain
invariant. However, the geodesic equations are invariant not only under
arbitrary transformation of coordinates (it is rather obvious) but also under 
 geodesic mappings $\Gamma_{\beta\gamma}^{\alpha}(x)\rightarrow\overline{\Gamma}_{\beta\gamma}^{\alpha}(x)$
of the Christoffel symbols in any fixed
coordinate system  \cite{Weyl}-\cite{Eisenhart}.

If $\Gamma_{\beta\gamma}^{\alpha}$ are Christoffel symbols in some coordinate
system $(x^{0}, x^{1}, x^{2}, x^{3}) $, and we use coordinate time $t=x^{0}/c$
($c$ is speed of light) as a  parameter along geodesic lines,
then the differential equations of  a geodesic line are of the form
\begin{equation}
\ddot{x}^{\alpha}+(\Gamma_{\beta\gamma}^{\alpha}-c^{-1}\Gamma_{\beta\gamma
}^{0}\dot{x}^{\alpha})\dot{x}^{\beta}\dot{x}^{\gamma}=0,
\label{EqMotionOfTestPart}%
\end{equation}
where $\dot{x}^{\alpha}=dx/dt,$ $\ddot{x}^{\alpha}=d\dot{x}^{\alpha}/dt$. It
easily to verify that these equations are invariant under the mapping
\begin{equation}
\label{GammaGeodesTransformations}\overline{\Gamma}_{\beta\gamma}^{\alpha
}(x)=\Gamma_{\beta\gamma}^{\alpha}(x)+\delta_{\beta}^{\alpha}\ \phi_{\gamma
}(x)+\delta_{\gamma}^{\alpha} \phi_{\beta}(x),
\end{equation}
where $\phi_{\alpha}(x)$ is  a continuously differentiable vector
field. In Riemannian space-time $\phi_{\alpha}(x)$
 can be expressed in terms of
determinants $g$ and $\overline{g}$ of the metric tensors before and after the
geodesic mapping of space-time $V$ into $\overline{V}$ as follows:%

\begin{equation}
\phi_{\alpha}=\frac{1}{n+1}\left(  \Gamma_{\alpha\gamma}^{\gamma}-\overline
{\Gamma}_{\alpha\gamma}^{\gamma}\right)  =\frac{1}{2(n+1)}\frac{\partial
}{\partial x^{\alpha}}\ln\left\vert \frac{\overline{g}}{g}\right\vert. 
\end{equation}

Such mappings of the Christoffel symbols in a given
coordinate system induce  some transformations not only  of the curvature and Ricci tensors  but also  transformations 
 $g_{\alpha\beta}\rightarrow\overline{g}_{\alpha\beta}$ of the metric tensor  which are obtained by solving of the partial differential equation
\begin{equation}
\overline{g}_{\beta\gamma;\alpha}(x)=2\phi_{\alpha}(x)\overline{g}%
_{\beta\gamma}(x)+\phi_{\beta}(x)\overline{g}_{\gamma\alpha}(x)+\phi_{\gamma
}(x)\overline{g}_{\alpha\beta}(x),
\label{gGeodesTransformation}%
\end{equation}
where the semicolon denotes a covariant derivative with respect to the metric in
$V_{4}$.

It is clearly that all solutions of these equation are, in principle,
equivalent physically. However  Einstein's equations even in vacuum are not invariant with respect
to geodesic mappings \cite{Petrov}. 

It follows from (\ref{GammaGeodesTransformations}) that the components $\overline{\Gamma}^{i}_{00}=\Gamma^{i}_{00}$. Therefore, in Newtonian limit 
geodesic-invariance is not an essential fact. However this cannot be said about
the relativistic case. For this reason is of interest to explore a theory in which the
gravitation equations, as well as the equations of motion of test particles, are
geodesic- invariant i.e. in which geodesic invariance plays the role of gauge
invariance.

In Sect. 2 a geodesic-invariant generalization of Einstein's equations is considered. It turns out that
 the simplest equations of such a kind are some bimetric equations.  In Sects˙ 3 to 7, proceeding   from 
new sight at  Poincar\'{e}  old fundamental ideas, we argue that the both spaces, Riemannian (with the curvature other than
 zero) and ﬂat,  can have physical sense depending on which kind of  reference frame is used.
In Sects. 8,  9 we find the spherically-symmetric solution of the above equations.
Finally, in Sect. 10 it is shown that features of the gravitational force resulting from
 this solution, well explain properties of the Hubble diagram obtained from the latest observations.

\section{Generalization of Einstein's equations}

The simplest way to find a gauge-invariant generalization of Einstein's
equations is to consider components  $\Gamma
_{\alpha\beta}^{\gamma}$  of the affine connection   in $n=4$ -dimensional manifold $\mathcal{M}_{4}$ as
4-components of some affine connection $\Gamma_{BC}^{A}$ of $n+1$ dimensionall
manifold $\mathcal{M}_{5}$  
in which transformations (\ref{GammaGeodesTransformations}) are 
consequences of 5th coordinate transformations \cite{Thomas}:%
\begin{equation}
\bar{x}^{4}=x^{4}-\int_{0}^{x}\phi_{\beta}(x^{\prime})dx^{\prime\beta},
\label{5-coordinate_transform}%
\end{equation}
where $\phi_{\beta}(x')$ is a gradient vector field. These transformations
together with  4-transformations
\begin{equation}
\bar{x}^{\alpha}=\bar{x}^{\alpha}(x^{0},x^{1},x^{2},x^{3})
\label{4transformations}%
\end{equation}
form some acceptable coordinate transformations in $\mathcal{M}_{5}$ . In
$\mathcal{M}_{5}$ one can  define an affine connection by the  functions
$\Gamma_{AB}^{A}$ which are transformed under (\ref{5-coordinate_transform})
and (\ref{4transformations}) as follows:
\begin{equation}
\overline{\Gamma}_{BC}^{A}=\left(  \Gamma_{ED}^{F}\frac{\partial x^{E}%
}{\partial\overline{x}^{B}}\frac{\partial x^{D}}{\partial\overline{x}^{C}%
}+\frac{\partial^{2}x^{F}}{\partial\overline{x}^{B}\partial\overline{x}^{C}%
}\right)  \frac{\partial\overline{x}^{A}}{\partial x^{F}}.
\label{5-KristoffelTransform}%
\end{equation}
The equations for 4-components:
\begin{equation}
\overline{\Gamma}_{\beta\gamma}^{\alpha}=\left(  \Gamma_{EC}^{F}\frac
{\partial\overline{x}^{E}}{\partial x^{\beta}}\frac{\partial x^{B}}%
{\partial\overline{x}^{\gamma}}+\frac{\partial^{2}x^{F}}{\partial\overline
{x}^{\beta}\partial\overline{x}^{\gamma}}\right)  \frac{\partial\overline
{x}^{\alpha}}{\partial x^{F}}%
\end{equation}
under transformations (\ref{5-coordinate_transform}) takes the form
(\ref{GammaGeodesTransformations}) if Christoffel's symbols satisfy the
conditions : $\Gamma_{\gamma4}^{\alpha}=\Gamma_{4\gamma}^{\alpha}%
=\delta_{\gamma}^{\alpha}$ and $\Gamma_{44}^{\alpha}=0$. These conditions are
invariant under coordinate transformations (\ref{4transformations}) and
(\ref{5-coordinate_transform}). The rest components of $\Gamma_{BC}^{A}$
should satisfy the conditions which also are obtained from consideration of
their transformation properties under 
(\ref{5-coordinate_transform}). The components $\overline{\Gamma}_{0\gamma
}^{4}=0$ in (\ref{5-KristoffelTransform}) do not depend on $\Gamma_{ED}^{F}$.
This allows to set $\Gamma_{0\gamma}^{4}=0$ in any coordinate system. The
component $\overline{\Gamma}_{44}^{4}$ are equal to ${\Gamma}_{44}^{4}$, which
allows to set ${\Gamma}_{44}^{4}=1$ in any coordinate system. The components
$\Gamma_{\alpha\beta}^{4}$ are transformed as a tensor under coordinate
transformations (\ref{4transformations}), and as
\begin{equation}
\overline{\Gamma}_{\alpha\beta}^{4}=\Gamma_{\alpha\beta}^{4}-\psi_{\alpha
\beta} \label{TransGamma5}%
\end{equation}
under transformations (\ref{5-coordinate_transform}) of the 5th coordinate
where $\psi_{\alpha\beta}=\psi_{\alpha;\beta}-\psi_{\alpha}\psi_{\beta},$ and
$\psi_{\alpha;\beta}$ is a covariant derivative in $V_{4}$.

The object
\begin{equation}
\mathcal{R}_{BCD}^{A}=\partial\Gamma_{BC}^{A}/\partial x^{D}\partial
\Gamma_{BD}^{A}/\partial x^{C}+\Gamma_{MD}^{A}\Gamma_{BC}^{M}-\Gamma_{MC}%
^{A}\Gamma_{BD}^{M}%
\end{equation}
is a curvature tensor in ${\mathcal{M}}_{5}$ . In our case the  components
$\mathcal{R}_{4\alpha}=\mathcal{R}_{\alpha4}$ and the component $\mathcal{R}_{44}$ of the
contracted tensor $\mathcal{R}_{BC}=\mathcal{R}_{BAC}^{A}$ are equal to zero
identically. For this reason $\mathcal{R}_{\beta\gamma}$ is the 4- tensor
\begin{equation}
{\mathcal{R}}_{\beta\gamma}=R_{\beta\gamma}+(n-1)\Gamma_{\beta\gamma}^{4},%
\end{equation}
where $R_{\beta\gamma}$ is the Ricci tensor.

The components of the Ricci tensor are transformed under
(\ref{5-coordinate_transform}) as follows:
\begin{equation}
\overline{R}_{\alpha\beta}=R_{\alpha\beta}-\psi_{\alpha\beta}.
\end{equation}
On account of (\ref{TransGamma5}) the object ${\mathcal{R}}_{\beta\gamma}$ is
gauge (i.e geodesic) invariant. For this reason, the simplest
geodesic-invariant generalization of  vacuum Einstein's equations is of the
form
\begin{equation}
R_{\beta\gamma}+(n-1)\Gamma_{\beta\gamma}^{4}=0, \label{NewRicci}%
\end{equation}
where $\Gamma_{\beta\gamma}^{4}$ is a tensor field which is transformed under
geodesic mappings according to (\ref{TransGamma5}).

Now the problem is to find some relevant  tensor $\Gamma_{\beta\gamma}^{4}$.
Such object must satisfy  the following requirements:

1. It should  transform under (\ref{5-coordinate_transform}) according to
(\ref{TransGamma5}),

2. It should transform under (\ref{4transformations}) as a 4-tensor,

3. It should lead to such generalization of Einstein's equations which solutions
can noticeably differ from the solutions of the last  equations only in
strong gravitation field as in  weak and moderately strong field solutions of
Einstein equations are in good agreement with observations.

The first requirement means that $\Gamma_{\beta\gamma}^{4}$ should be formed
from geometrical objects of space-time $V_{4},$ and cannot be an additional
field. 
Because the equations of motion of free particles contains   connection coefficients
(and not metric) it should be formed from components of 
$\Gamma_{\beta\gamma}^{\alpha}$.

 The  simplest object  $\Gamma_{\beta\gamma}^{4}$   which can be formed by components of
Christoffel symbols and which has  the requirement transformations
properties under our gauge transformations is
\begin{equation}
\Gamma_{\alpha\beta}^{4}=\frac{1}{n+1}(\Gamma_{\alpha,\beta}-\Gamma
_{\alpha\beta}^{\gamma}\ \Gamma_{\gamma})-\frac{1}{(n+1)^{2}}\Gamma_{\alpha
} \Gamma_{\beta},
 \label{Gamma0Temp}%
\end{equation}
where $\Gamma_{\alpha}=\Gamma_{\alpha\gamma}^{\gamma}.$ In this case
${\mathcal{R}}_{\beta\gamma}$ is an object which is formed by gauge-invariant
Thomas symbols \cite{Thomas}
\begin{equation}
\Pi_{\alpha\beta}^{\gamma}=\Gamma_{\alpha\beta}^{\gamma}-(n+1)^{-1}\left[
\delta_{\alpha}^{\gamma}\ \Gamma_{\beta}+\delta_{\beta}^{\gamma}%
\ \Gamma_{\alpha}\right]  \; \label{ThomasSymbols},%
\end{equation}
the same way as the Ricci tensor is formed out of the Christoffel symbols. As $\Pi_{\alpha\gamma
}^{\alpha}=0$, equations (\ref{NewRicci})  take the form
\begin{equation}
\nabla_{\gamma}\Pi_{\beta}^{\alpha}-\Pi_{\beta\delta}^{\epsilon}\Pi_{\epsilon\gamma
}^{\delta}=0 \label{MyEqsByThomases}%
\end{equation}
where $\nabla_{\alpha}$ denotes a covariant derivative in Minkowski space-time.

However, the problem is that ${\mathcal{R}}_{\beta\gamma}$ is not a tensor with respect to general
coordinate transformations because $\Gamma_{\alpha}(x)$ is not a vector field.

This a new problem evidently cannot be solved within the framework of General
Relativity because by using components $\Gamma_{\beta\gamma}^{\alpha}$ it is
impossible to form a tensor of the second rank in $V_{4}$ which is invariant
under geodesic transformations.

However  a satisfactory solution of this problem can be found if
the Riemannian space-time $V_{4}$ coexists in the manifold $\mathcal{M}_{4}$ with  Minkowski
space-time $E_{4}$ which also has a physical meaning. From the
point of view of such bimetric structure of space-time, in Minkowski
space-time $g_{\alpha\beta}(x)$ is simpy some tensor field. The Christoffel symbols
$\Gamma_{\alpha\beta}^{\gamma}$ can be considered as  components of the tensor $D_{\beta\gamma
}^{\alpha}=\Gamma_{\beta\gamma}^{\alpha}-\overset{\circ}{\Gamma}_{\beta\gamma
}^{\alpha}$ in Cartesian coordinates, where $\overset{\circ}{\Gamma}_{\beta\gamma
}^{\alpha}  $
is the Christoffel symbols in Minkowski space-time. Tensor $D_{\beta\gamma
}^{\alpha}$ is formed by substitution of ordinary derivatives in $\Gamma_{\alpha\beta
}^{\gamma}$ by the covariant ones \footnote{There is an analogy here with
the formalism of the bimetric Rosen theory \cite{Treder}.} because both the
definitions of the tensor coincide in Cartesian coordinates. The object
$\Gamma_{\alpha}$ is components of the vector $D_{\alpha}=D_{\alpha\beta}^{\beta}%
=\Gamma_{\alpha\beta}^{\beta}-\overset{\circ}{\Gamma^{\beta}}_{\alpha\beta}$
in Cartesian coordinates.

In such bimetric manifold we can set%
\begin{equation}
\Gamma_{\alpha\beta}^{4}=Q_{\alpha,\beta}-\Gamma_{\alpha\beta}^{\gamma
}\ Q_{\gamma}-Q_{\alpha}\ Q_{\beta}=Q_{\alpha;\beta}-Q_{\alpha}\ Q_{\beta},
\label{GammaFinal}%
\end{equation}
where%
\begin{equation}
Q_{\alpha}= \frac{1}{n+1} D_{\alpha}=\frac{1}{2(n+1)}\frac{\partial}{\partial
x^{\alpha}}\ln\left\vert \frac{g}{\eta}\right\vert .
\end{equation}
The definition (\ref{Gamma0Temp}) is  the tensor
(\ref{GammaFinal}) in Cartesian coordinates.

Similarly to it, instead of Thomas's symbols (\ref{ThomasSymbols}) we have the gauge-invariant 
 tensor
\begin{equation}
B_{\alpha\beta}^{\gamma}=D_{\alpha\beta}^{\gamma}-\left[  \delta_{\alpha
}^{\gamma}\ Q_{\beta}+\delta_{\beta}^{\gamma}\ Q_{\alpha}\right]  \;
\end{equation}
It can be written also as%
\begin{equation}
B_{\beta\gamma}^{\alpha}=\Pi_{\beta\gamma}^{\alpha}-\overset{\circ}{\Pi
}_{\beta\gamma}^{\alpha}%
\end{equation}
where
\begin{equation}
\overset{\circ}{\Pi}_{\alpha\beta}^{\gamma}=\overset{\circ}{\Gamma}%
_{\alpha\beta}^{\gamma}-(n+1)^{-1}\left[  \delta_{\alpha}^{\gamma}%
\overset{\circ}{\Gamma}_{\beta}+\delta_{\beta}^{\gamma}\overset{\circ}{\Gamma
}_{\alpha}\right]  \;
\end{equation}
are Thomas symbols for $E_{4}$ in the used coordinate system.

Therefore, Eqs. (\ref{MyEqsByThomases}) can be considered as a particular case of the more general equations%
\begin{equation}
\nabla_{\alpha}B_{\beta\gamma}^{\alpha}-B_{\beta\delta}^{\epsilon}%
B_{\epsilon\gamma}^{\delta}=0, \label{MyVacuumEqs}%
\end{equation}
where $\nabla_{\alpha}$ denotes a covariant derivative in $E_{4}$.

The above can be generalized to the case of matter presence. It is naturally to consider 
the tensor $g_{\alpha\beta}$ as 4-components of
some metric tensor $g_{AB}$ in the manifold $\mathcal{M}_{5}:$%
\begin{equation}
g_{AB}=\left(
\begin{array}
[c]{cc}%
g_{\alpha\beta} & g_{\alpha4}\\
g_{4\alpha} & g_{44}%
\end{array}
\right)  .
\end{equation}
The components $g_{\alpha\beta}$ are transformed under
(\ref{5-coordinate_transform}) as follows:%
\begin{align}
\overline{g}_{\alpha\beta}  &  =g_{\alpha\beta}+g_{44}\phi_{\alpha}\phi
_{\beta}+g_{4\alpha}\phi_{\beta}+g_{4\beta}\phi_{\alpha}%
\label{gABtransformations}\\
\overline{g}_{\alpha4}  &  =g_{\alpha4}+g_{44}\phi_{\alpha}\\
\overline{g}_{\alpha4}  &  =g_{44}.
\end{align}
Transformations of the components $\Gamma_{\alpha}$ under geodesic mappings
are given by%
\begin{equation}
\overline{\Gamma}_{\alpha}=\Gamma_{\alpha}+(n+1)\phi_{\alpha},
\end{equation}
which coincides with the transformations of the components $g_{\alpha4}$ if
$g_{44}=n+1$.  For this reason we can assume that \footnote{It means that there is some relationship between
$\Gamma^{A}_{BC} $ and  $g_{AB}$ distinct from such relationship in Riemannian geometry.}
\begin{equation}
g_{AB}=\left(
\begin{array}
[c]{cc}%
g_{\alpha\beta} & Q_{\alpha}\\
Q_{\alpha} & n+1
\end{array}
\right)  .
\end{equation}
Then there is a gauge-invariant tensor%
\begin{equation}
\hat{g}_{\alpha\beta}=g_{\alpha\beta}-(n+1)^{-1}Q_{\alpha}Q_{\beta}.
\end{equation}

Thus, finally, the gauge-invariant generalization of  Einstein's  equations
with matter are of the form%
\begin{equation}
\nabla_{\alpha}B_{\beta\gamma}^{\alpha}-B_{\beta\delta}^{\epsilon}%
B_{\epsilon\gamma}^{\delta}=k\left(  T_{\alpha\beta}-\frac{1}{2}\hat{g}_{\alpha
\beta}T\right)  , \label{myEqsWithMatter}%
\end{equation}
where $k=8\pi\gamma/c^{4},$ $\gamma$ is the gravitational constant, $c$ is the
speed of light, $T_{\alpha\beta}$ is the matter energy-momentum tensor. At the
gauge conditions $Q_{\alpha}=0$ equations (\ref{myEqsWithMatter}) coincides with
 Einstein's equations.

Of course, these bimetric equations may be true if and only if both the space-times,
$V_{4}$ and $E_{4}$, has some physical meaning.
How these two physical space-time can coexist?
Attempts to use flat space-time simultaneously with curved in general
relativity were undertaken repeatedly. Here it should be mentioned the well
known Rosen's bimetric theory \cite{Rosen} which leads to non-Einstein
equations for $g_{\alpha\beta}(x)$ , and interpretations of the Einstein's
equations as the equations for $g_{\alpha\beta}$ considered as some function
of a tensor field $\psi_{\alpha\beta}$ which describes gravitational field in
Minkowski space-time \cite{Thirring}, \cite{Deser}, \cite{Ogievetski},
\cite{Zeldovich}. The need to go beyond Riemannian manifold has noted been also 
in \cite{Hehl}.

It should be noted that at any concrete expressions of $g_{\alpha\beta}$ in
terms of $\psi_{\alpha\beta},$ which are used in some papers, the Einstein
equations, considered as a function of $\psi_{\alpha\beta},$ cannot be
invariant with respect to possible gauge transformations of the field
$\psi_{\alpha\beta}$. In the paper \cite{Verozub91} it was shown that the
requirements of the invariance of the any equations for $g_{\alpha\beta}$ with
respect to the possible gauge transformations of $\psi_{\alpha\beta}$ leads
 uniquely to geodesic invariance of gravitational equation, and the
simplest generalization of Einstein's vacuum equations are 
equations (\ref{NewRicci}) as obtained above.

However bimetric nature of physical space-time 
 belongs probably to one of the 
deepest problems of physics. In Sec. 3 to 7 we show how 
both the kind of space-time can coexist in the theory.

\section{Relativity of space-time}

At the beginning of the 20th century H. Poincar\'{e} convincingly showed that
it makes no sense to assert that one or other geometry of physical space is
true. Only an aggregate " geometry + measuring instruments" has a physical,
 meaning verifiable by experiment. Einstein recognized that Poincar\'{e} reasoning
is justified. His theory of gravity does not reject Poincar\'{e}'s
ideas. It only demonstrates relativity of space and time geometry with respect
to distribution of matter. Poincar\'{e}'s ideas indicate to another property
of physical reality --- relativity of the geometry in relation to measuring
instruments, dependency on their properties. These ideas, apparently, have
never been realized in physics and, evidently, remain until now a subject for
discussion of philosophers. The success of General Relativity almost convinced
us that physical space-time in the presence of matter is Riemannian and that
this fact does not depend on properties of measuring instruments. Attempts to
describe gravity in flat space-time have not obtained recognition and
Poincar\'{e}'s ideas have proved to be almost forgotten for physics.

However, are Poincar\'{e}'s ideas really only a thing of such little use as
conventionalism, as it is usually believed? In fact, a choice of certain
properties of measuring instruments is nothing more than the choice of some
frame of reference, which is just such a physical device by means of which we
test properties of space-time. (Of course, we must understand the distinction
between a frame of reference as a physical device, and a coordinate system as
only a means of parameterization of events in space-time.) For this reason
Poincar\'{e}'s reasoning about interdependency between properties of
space-time and measuring instruments should be understood as the existence of
a connection between geometry of space-time and properties  of the
employed reference frame .

The existence of such a connection can be shown by an analysis of a simple and
well-known example considering it from a new point of view. Disregarding the
rotation of the Earth, a reference frame, rigidly connected with the Earth
surface, can be considered as an inertial frame (IFR). An observer, located in
this frame, can describe the motion of freely falling identical point masses as
taking place in Minkowski space-time under the action of a certain force
field. However, consider another observer, who is located in a frame of
reference, the reference body of which is formed by these freely falling
particles. Such a reference frame can be named the proper frame of reference
(PFR) for the given force field. Let us assume that the observer is deprived
of the possibility of seeing the Earth and stars. This observer does not feel
the presence of the force field in any place of his frame. Therefore, if he
proceeds from the relativity of space-time in the
Berkeley-Leibnitz-Mach-Poincar\'{e} (BLMP) sense \cite{Berkeley}
-\cite{Poincare}, then from his viewpoint accelerations of the point masses,
forming the reference body of his frame, in his physical space must be equal
to zero. However, instead of this, he observes a change in distances between
these point masses in time. How can he explain this fact? Evidently, the only
reasonable explanation for him is the interpretation of this observed
phenomenon as a manifestation of the deviation of geodesic lines in some
Riemannian space-time of a nonzero curvature. Thus, if the first observer,
located in the IFR, can postulate that space-time is flat, the second
observer, located in a PFR of the force field, who proceeds from relativity of
space and time in the BLMP sense, already in the Newtonian approximation
\textit{is forced} to consider space-time as Riemannian with curvature other
than zero.

In next Section it is shown that a similar dual description is possible also for the motion of a perfect isentropic fluid.

\section{Space-time metric form in NIFRs}

It is possible to arrive at some quantitative relationship between properties
of the used frames of reference and the local metric properties of space-time in PRFs
examining the problem from more general positions.

 Presently we do not know how the space-time geometry in an inertial frame of
reference (IFR) is related with the frame properties. Under the circumstances
it can simply be postulated, according to Special Relativity, that space-time
in IFRs is pseudo-Euclidean. Proceeding from it, one can find the line element of space-time  in non-inertial frames of reference (NIFR) from the
viewpoint of observers located in the NIFRs and proceeded from relativity of
space and time in the BLMP meaning.

By a non-inertial frame of reference we mean the frame, the body of reference
of which is formed by point masses moving in an IFR under the effect of a
given force field.

The reference body (RB) of a reference frame is supposed to be formed by identical
 point masses $m$. If an observer in the reference frame   is at rest , his world
line coincides with the world line of some point of the reference body. It is
obvious for such an observer in an IFR that the accelerations of the point
masses forming his reference body are equal to zero.  This fact
takes place in relativistic meaning, too. That is, if the line element 
 of space-time in the IFR is denoted by $d\sigma$ and $u^{\alpha
}=dx^{\alpha}/d\sigma$ is the field  of 4- velocity of the
 point masses
forming the reference body, then the absolute derivative of 
$u^{\alpha}$ is equal to zero: \footnote{We use notations and
definitions, following the Landau and Lifshitz book \cite{Landau}.}
\begin{equation}
Du^{\alpha}/d\sigma=0. \label{DuDsigma}%
\end{equation}
(We mean that an arbitrary coordinate system is used.)

Does it occur for an observer in NIFRs ? That is, if the differential metric
form of space-time in a NIFR is denoted by $ds$, does the 4-velocity vector
$\zeta^{\alpha}=dx^{\alpha}/ds$ of the point-masses forming the reference body
of the NIFR satisfy the equation
\begin{equation}
D\zeta^{\alpha}/ds=0\, \, ?
\label{DzetaDs}%
\end{equation}
 If the space-time is absolute, equation (\ref{DzetaDs}) holds for only
$ds = d\sigma$. However, if space and time are relative in the BLMP sense, then
for both observers, located in some IFR and NIFR, the motion of the point masses,
forming their reference bodies, which are kinematically equivalent, must be
dynamically equivalent, too (both in non-relativistic and relativistic sense).
Any observer in the NIFR, isolated from the external world and proceeded from
relativity of space-time in BLMP meaning, consider points of the reference
body as the ones of his physical space, and space of events as his space-time.
Therefore, from his viewpoint point masses forming the reference body of his
frame are not under action of any forces (the same as for the observer in
IFR), and their 4-velocity must be equal to zero. In other words, since for
the observer in the IFR world lines of the reference body
are, according to  (\ref{DuDsigma}), some geodesic lines, for the observer
in the NIFR the world lines of the  of his BR also must be geodesic
lines in his space-time, which can be expressed by  (\ref{DzetaDs} ).

The equation (\ref{DzetaDs}) uniquely determines the fundamental metric form
in NIFRs. Indeed, the differential equations of these world lines are at the
same time Lagrange equations describing in Minkowski space-time the motion of
the point masses forming the reference bodies of the NIFR. The last equations
can be obtained from a Lagrange action $S$ by the principle of the least
action. Therefore, the equations of the geodesic lines can be obtained from a
differential metric form $ds=k\ dS$, where $k$ is  constant, $dS=\mathcal{
L}(x,\dot{x})dt$, and $\mathcal{L}(x,\dot{x})$ is a Lagrange function
describing in Minkowski space-time the motion of identical point masses $m$
forming the body reference of the NIFR. . The constant $k$ is equal to
$-(mc)^{-1}$, as it follows from the analysis of the case when the frame of
reference is inertial.

Thus, if we proceed from relativity of space and time in the BLMP sense, then
the line element of space-time in NIFRs can be expected to have
the following form
\begin{equation}
ds=-(mc)^{-1}\;dS(x,dx). \label{dsMain}%
\end{equation}

Therefore, properties of space-time in NIFRs are entirely determined by
properties of used frames in accordance with the BLMP idea of relativity of
space and time, which has been noted for the first time in  \cite{Verozub81a}.

Consider two examples of NIFRs.

1. The reference body is formed by noninteracting electric charges, moving in
a constant homogeneous electric field $\mathcal{E}$. The motion of the charges
in an IFR is described in the Cartesian coordinates system by a Lagrangian
\cite{Landau}
\begin{equation}
L=-m\,c^{2}\ (1-v^{2}/c^{2})^{1/2}+\mathcal{E}\,e\,x, \label{lagrframe1}%
\end{equation}
where $v$ is the speed of a particle. According to  (\ref{dsMain}) the
space-time metric differential form in this frame is given by%
\begin{equation}
ds=d\sigma-(wx/c^{2})dx^{0},
\end{equation}
where
\[
d\sigma=[\eta_{\alpha\beta}dx^{\alpha}dx^{\beta}]^{1/2}%
\]
is the differential metric form of the Minkowski space-time in the IFR in the
coordinate system being used, and $w=e\,\mathcal{E}/m$ is the acceleration of
the charges.

2. The reference body consists of noninteracting electric charges in a
constant homogeneous magnetic field $H$ directed along the axis $z$. The
Lagrangian describing the motion of the particles can be written as follows
\cite{Landau}:
\begin{equation}
L=-mc^{2}(1-v^{2}/c^{2})^{1/2}-(m\Omega_{0}/2)(\dot{x}y-x\dot{y}),
\label{LagrangGarge}%
\end{equation}
where $\dot{x}=dx/dt$, $\dot{y}=dx/dt$, and $\Omega_{0}=eH/2mc$.

The points of such a system rotate in the plane $xy$ around the axis $z$ with
the angular frequency
\begin{equation}
\omega=\Omega_{0}[1+(\Omega_{0}r/c)^{2}]^{-1/2} \label{omega (r)},%
\end{equation}
where $r=(x^{2}+y^{2})^{1/2}$. The linear velocity of the BR points tends to
$c$ when $r\rightarrow\infty$.

For the given NIFR
\begin{equation}
\label{dsInRotationNIFR}ds = d\sigma+ (\Omega_{0} / 2c)\ (ydx - xdy).
\end{equation}

In the above NIFRs $ds$ is of the form
\begin{equation}
ds=\mathcal{F}(x,dx), \label{dsRanders}%
\end{equation}
where $\mathcal{F}(x,dx)=d\sigma+f_{\alpha}(x)dx^{\alpha}$, $f_{\alpha}$ is a
vector field. Therefore, $\mathcal{F}(x,dx)$ is a homogeneous function of the first
degree in $dx^{\alpha}$, that is $\mathcal{F}(x,k y)=k\,\mathcal{F}(x,y)$. 
Thus, the space-time in the above NIFRs is Finslerian  \cite{Rund}. 

By using the identities
\begin{equation}
\mathcal{F}(x,y)=\frac{\partial{F(x,y)}}{\partial{y^{\alpha}}}\,\xi^{\alpha
};\,\, \, \quad\frac{\partial^{2}\mathcal{F}(x,y)}{\partial{y^{\alpha}}
\partial{y^{\beta}}}=0,
\end{equation}
the function $\mathcal{F}^{2}(x,y)$ can be written as%
\begin{equation}
\mathcal{F}(x,y)^{2}=G_{\alpha\beta}(x,y)\,y^{\alpha}y^{\beta},
\end{equation}
where
\begin{equation}
G_{\alpha\beta}(x,y)=\frac{1}{2}\frac{\partial^{2}\mathcal{F}(x,y)^{2}}{
\partial{y}^{\alpha}\partial{y}^{\beta}}%
\end{equation}
is an analog of the fundamentall metric tensor of Riemannian space-time.
Consequently, the modulus of the vector $y^{\alpha}$ in the point $x^{\alpha}$
is $||y||= \mathcal{F}(x,y)= [ G_{\alpha\beta}(x,y)y^{\alpha}y^{\beta}]^{1/2}$.

For $ds$ of the form (\ref{dsRanders}) we have
\begin{equation}%
G_{\alpha\beta}(x,y)= \eta_{\alpha\beta}+f_{\alpha}f_{\beta}
+(\eta_{\alpha\beta}- \tau_{\alpha}\tau_{\beta} ) f_{\sigma} \tau^{\sigma
}+f_{\alpha}\tau_{\beta} + f_{\beta}\tau_{\alpha},%
\end{equation}
where $\tau^{\alpha}=dx^{\alpha}/d\sigma$.

A covariant vector in the Finslerian space-time, therefore, can be defined as%
\begin{equation}
\overset{\ast}{y}_{\alpha}= \mathcal{F}(x,y)\frac{\partial{\mathcal{F}(x,y)}%
}{\partial{y^{\alpha}}} = G_{\alpha\beta}(x,y)\,y^{\alpha}.%
\end{equation}

The orthogonality of vectors $y^{\alpha}$ and $y_{1}^{\alpha}$ can be defined
by the equality $\overset{\ast}{y}_{\alpha} y_{1}^{\alpha}=0$. It must be
noted that the orthogonality of two vectors is not symmetric.

3.   The reference body is formed by particles of an isentropic  flow of the
perfect fluid. 

It is well known that besides of traditional continual description,  a perfect  fluid can be considered as a collection of
a finite number of identical macroscopic small particles which are  under influence of interparticles forces 
which mimic the effect of pressure, viscosity  etc   \cite{Monaghan0} , \cite{Monaghan2}.
In particular, the fluid velocity in a given point is simply velocity of the particle  being in this point. 
At such description the motion of the fluid particles is governed
by solutions of ordinary  differential equations of Newtonian or relativistic dynamics. 
 
 The motion of the
fluid particles in an IFR is described by  Lagrangian \cite{Verozub07a}%
\begin{equation}
L=-mc \left(  G_{\alpha\beta}\frac{dx^{\alpha}}{d\lambda}\frac{dx^{\beta}%
}{d\lambda}\right)  ^{1/2}d\lambda\label{Lagrangian_in_V},%
\end{equation}
where $\lambda$ is a parameter along 4-path of particles, 
$w$  is   enthalpy per unit volume, 
$G_{\alpha\beta
}=\varkappa^{2}\eta_{\alpha\beta}$ , $\varkappa=w/\rho c^{2}$, $\rho=m n$, 
$m$ is the mass of the particles,
$n$ is  the
particles number density,  and
$\eta_{\alpha\beta}$ is the metric tensor in Minkowski space-time. According
to  (\ref{dsMain}) the line element of space-time  in this frame is
given by
\begin{equation}
ds^{2}=G_{\alpha\beta}dx^{\alpha}dx^{\beta} . \label{ds2}%
\end{equation}
The covariant derivative of tensor $G_{\alpha\beta}$ is equal to zero.
Therefore, space-time in such NIFR is Riemannian with the curvature other than zero. 
The last description  is equivalent  to
the one by the Lagrangian (\ref{Lagrangian_in_V})  \cite{Verozub07a}.  

  Thus,  an observer in some laboratory
(inertial) frame of reference can describe a fluid motion as what is going on in
Minkowski space-time by a Lagrangian. However, the observer which is
located in a comoving frame of reference cannot feel presence of the force
field arising from a pressure gradient since he moves along a geodesic line of
some Riemannian space-time. Therefore, if he is isolated from external world,
he is forced to explain relative motions of nearby particles of the fluid as a
manifestation of deviation of geodesic lines due to space-time curvature in this frame of reference.


\section{3+1 decomposition of space-time in NIFRs and Sagnac effect}

For the $3+1$ decomposition of space-time in non-inertial frames of
reference into physical 3-space and time we use a covariant method which is a
Finslerian generalization of the well-known method in General Relativity
\cite{Dehnen}, \cite{Efremov}.

An ideal clock is a local periodic process measuring the length of its own
world line on a certain scale. For an observer in a NIFR the direction of
physical time in a point $x^{\alpha}$ is given by the vector of the 4-velocity
$\zeta^{\alpha}=dx^{\alpha}/ds$ of the given point of the NIFR reference body.

By analogy with the case of Riemannian geometry, an arbitrary 4-vector
$\xi^{\alpha}$ in the point $x^{\alpha}$ can be represented as follows:%
\begin{equation}
\xi^{\alpha}=\overline{\xi}^{\alpha}+\beta\zeta^{\alpha}. \label{decomp4vect}%
\end{equation}
In this equation $\beta$ is a function of $x^{\alpha}$, and $\overline{\xi
}^{\alpha}$ is a 3-spatial component of the vector $\xi^{\alpha}$. 
It is supposed to be satisfied to
Finslerian's orthogonality condition \cite{Rund} to the time direction which
we define as%
\begin{equation}
\overset{\ast}{\zeta}_{\alpha}\overline{\xi}^{\alpha}=0, \label{ortogonality}%
\end{equation}
where $\overset{\ast}{\zeta}_{\alpha}$ is the covariant components of the
vector $\zeta^{\alpha}$:
\begin{equation}
\label{covarvect}\overset{\ast}{\zeta}_{\alpha}=\mathcal{F}(x,\zeta
)\frac{\partial{\mathcal{F }(x,\zeta)}}{\partial{\zeta^{\alpha}}}.%
\end{equation}

Since $\mathcal{F}(x,\zeta)=1$, this vector is
\begin{equation}
\overset{\ast}{\zeta_{\alpha}}=\frac{\eta_{\alpha\beta}\zeta^{\beta}}{\left(
\eta_{\alpha\beta}\zeta^{\alpha}\zeta^{\beta}\right)  ^{1/2}} +f_{\alpha}.
\label{decompositioncovectora}%
\end{equation}
By multiplying    (\ref{decomp4vect}) with the vector $\overset{\ast}{
\zeta_{\alpha}}$ we find that $\beta=\overset{\ast}{\zeta} _{\alpha}%
\xi^{\alpha}$ and
\begin{equation}
\overline{\xi}^{\alpha}=H_{\beta}^{\alpha}\xi^{\beta}, \label{spacecomp}%
\end{equation}
where $H_{\beta}^{\alpha}=\zeta^{\alpha}\overset{\ast}{\zeta} _{\beta}%
-\delta_{\beta}^{\alpha}$ is an operator of spatial projection, and
$\delta_{\beta}^{\alpha}$ is the Kronecker delta.

For the vector $\xi^{\alpha}=dx^{\alpha}$  (\ref{decomp4vect}) yields
\begin{equation}
dx^{\alpha}=\overline{dx}^{\alpha}+c\ dT_{f}\zeta^{\alpha}, \label{3+1For_dx}%
\end{equation}
where $\overline{dx}^{\alpha}$ is the spatial components of the vector
$dx^{\alpha},$ and
\begin{equation}
dT_{f}=c^{-1}\overset{\ast}{\zeta}_{\alpha}dx^{\alpha} \label{dT}%
\end{equation}
is the time element between events in the points $x^{\alpha}$ and $x^{\alpha
}+dx^{\alpha}$ in the NIFR.

Metric form (\ref{dsRanders}) and the spatial projection of the vector
$dx^{\alpha}$ leads to the following covariant form of the spatial element in
NIFRs
\begin{equation}
dL_{f}=(-\eta_{\alpha\beta}\overline{dx}^{\alpha}\overline{dx} ^{\beta}%
)^{1/2}+f_{\alpha}\ \overline{dx}^{\alpha} \label{SpatialElemFisler}.%
\end{equation}

This covariant equation is the simplest and clearest in the "comoving"
coordinate systems, in which
\begin{equation}
\label{ComovingZeta}\zeta^{\alpha}= \varepsilon\delta_{0}^{\alpha}.%
\end{equation}
It follows from the equality $\mathcal{F}(x,\zeta)=1$ that $\varepsilon
=(\eta_{00}^{1/2}+f_{0})^{-1}$.

In this coordinate system for an observer at rest the time element $dT_{f}%
=(\eta_{00}^{1/2}+f_{0} ) dx^{0}$, $\overline{dx }^{i}=dx^{i}$, and
$\overline{dx}^{0}=0$. Therefore,
\begin{equation}
dL_{f}=(-\eta_{ik}dx^{i}dx^{k})^{1/2}+f_{i}dx^{i}=dL(1+f_{i}k^{i}),
\label{DLaprox}%
\end{equation}
where $dL=(-\eta_{ik}dx^{i}dx^{k})^{1/2}$ is the Euclidean spatial element and
$k^{i}=dx^{i}/dl$ is the unit direction vector.

Of course, the above definitions of $dT_{f}$ and $dL_{f}$ have physical
meaning only in the space-time region where they are real and positive
quantities. In this respect, space-time in NIFRs has more complicated
structure as compared with IFRs, and depends on properties of the considered
NIFR. For example, space-time in the first NIFR in Sec. 4 is time-like only
if the interval $d\sigma$ is time-like and the distance $|x|$ of charges from
the coordinate origin satisfy the condition $|x| < c^{2}/ w $.

The Finslerian geometry of space-time in NIFRs allows us to give a natural
explanation of the well known Sagnac effect, i.e. of a phase shift in the interference of
two oppositely directed
coherent light beams on a rotating disk \cite{Post}. It is usually, for
relativistic explanation of this effect the motion of light in a NIFR is considered as
a relative motion in the
Pseudo-Euclidean space-time of some IFR \cite{Ashtekar}. However, for an 
 observer located in a rotating frame and isolated from the IFR (in a
\textquotedblright black box\textquotedblright) , who proceeds from the notion
of space and time relativity in BLMP sense , the observed anisotropy in time
of light propagation clockwise and counterwise
is not a trivial effect. It must have some \textquotedblright
internal\textquotedblright\ physical explanation.

Consider a rigid disk rotating in the plane $xy$ with a constant angular
velocity $\Omega$ around the axis $z$. Let $r$ and $\theta$ be the
coordinates, defined by the equations
\begin{equation}
x=r\cos\varphi,\; y=r\sin\varphi,\; \varphi=\theta+\Omega t. \label{7/-}%
\end{equation}
In the coordinate system $(r,\theta,z,t)$ the differential metrical form $ds$
is of the form
\begin{equation}
ds=d\sigma-(\Omega r^{2}/2c)\, d\theta-(\Omega^{2}r^{2}/2c^{2}) dx^{0},
\label{7/10}%
\end{equation}
where $d\sigma$ is the pseudo - Euclidean metric form:
\begin{equation}
\label{detainsphercoord}d\sigma^{2} =(1-\Omega^{2} r^{2}/c^{2})\, {dx^{0}}^{2}
- dr^{2}-r^{2}d\theta^{2}
-2(r^{2}\Omega/c)d\theta\,dx^{0}-dz^{2}.\;\;
\end{equation}

At $\Omega R/c \ll1$ this reference frame  is identical to the NIFR described
in example 2 of Sec. 4, and the used coordinate system on the disk is
"comoving", where  (\ref{ComovingZeta}) holds.

It follows from  (\ref{DLaprox}) that the spatial element in the NIFR is
anisotropic. We will show that the speed of light in the noninertial frame of
reference is anisotropic, too. Along photon paths the equality $\eta
_{\alpha\beta} dx^{\alpha} dx^{\beta}=0 $ holds. By using 
(\ref{3+1For_dx}) , this equation can be written in terms of notions of the
NIFR as
\begin{equation}
\eta_{\alpha\beta} \overline{dx}^{\alpha} \overline{dx}^{\beta}+2\, c\,
dT_{f}\, \eta_{\alpha\beta}\zeta^{\alpha} \overline{dx}^{\beta}
+c^{2} dT_{f}^{2} \eta_{\alpha\beta} \zeta^{\alpha} \zeta^{\beta}=0,
\label{EqForPhoton0}%
\end{equation}
where the 4-vector $\overline{dx}^{\alpha}$ is orthogonal to $\zeta^{\alpha}$
, i.e.
\begin{equation}
\left[  \frac{\eta_{\alpha\beta}\zeta^{\beta}}{\left(  \eta_{\alpha\beta}%
\zeta^{\alpha}\zeta^{\beta}\right)  ^{1/2}}+f_{\alpha}\right]  \overline
{dx}^{\alpha} =0.
\end{equation}
For this reason the preceding equation takes the form
\begin{equation}
\eta_{\alpha\beta}\,v_{ph}^{\alpha}v_{ph}^{\beta} -2\, c\, f_{\alpha}
v_{ph}^{\alpha} (\eta_{\alpha\beta}\zeta^{\alpha} \zeta^{\beta})^{1/2}%
+c^{2}\eta_{\alpha\beta}\zeta^{\alpha}\zeta^{\beta}=0,
\label{exactEqForLightVelocityOnDisk}
\end{equation}
where $v_{ph}^{\alpha}= \overline{dx}^{\alpha}/dT_{f}$ is a photon velocity as
measured by an observer in the NIFR. 

The covariant equation (\ref{exactEqForLightVelocityOnDisk})
become more
clearly in the coordinate system where the conditions (\ref{ComovingZeta}) are satisfied.

Indeed, it follows from the equality $\mathcal{F}(x,\zeta)=1$ and from
(\ref{ComovingZeta} ) and (\ref{7/10}) that the quantity $(\eta_{\alpha\beta
}\zeta^{\alpha}\zeta^{\beta})^{1/2}=1- f_{\alpha}\zeta^{\alpha}=1- f_{2}%
\zeta^{2}- f_{0}\zeta^{0}=1$ with accuracy up to $\Omega r/c$. Therefore,
photon's velocity $v_{ph}$ in the NIFR satisfy the equation
\begin{equation}
v_{ph}^{2}+2 c\, v_{ph}\, f_{i} k^{i}-c^{2}=0,
\end{equation}
where
\begin{equation}
v_{ph}^{2}=-\eta_{ik}\frac{ dx^{i}}{dT_{f}} \frac{dx^{k}}{dT_{f}}.
\end{equation}
Thus, the velocity of an outgoing photon is given up to accuracy $\Omega r/c$
by the equation
\begin{equation}
v_{ph}=c\, ( 1-f_{i}k^{i} ). \label{vNIFR}%
\end{equation}

The time of the photon motion around of the disk of the radius $R$ is an
integral of
\begin{equation}
dT_{f}=\frac{dL_{f}}{v_{ph}}=\frac{dL}{c}(1+2 f_{i}k^{i}),
\end{equation}
which yields for clockwise and counterclockwise directions $dT_{F+}=(dL/c)
(1+2 f_{\theta})$ and $dT_{F-}=(dL/c) (1-2 f_{\theta})$, respectively. The
difference in the time interval between light propagation on the rotating disk
in two directions is $4\pi R^{2}\Omega/c^{2}$, which gives the Signac phase
shift \cite{Post}. Thus, the Signac effect for an isolated observer in the
rotating frame can be treated as caused by the Finslerian metric of space-time
in non-inertial frames of reference, see also \cite{Verozub81b}.

\section{ Inertial Forces}

Let us show that the existence of inertial forces in NIFRs can be interpreted
as a manifestation of a Finslerian connection of space-time in these frames.

According to our main assumption in Sec. 2, the differential equations of
the motion of the point masses in an IFR , forming the reference body of a
NIFR, are equations of geodesic lines of space-time in this NIFR. These
equations can be found from the variational principle $\delta\int ds=0$ and
are of the form
\begin{equation}
d\zeta^{\alpha}/ds+G^{\alpha}(x,\zeta)=0, \label{11/29}%
\end{equation}
where $\zeta^{\alpha}$ is the 4-velocity of the point mass, the world line of
which is $x^{\alpha}=x^{\alpha}(s)$, and
\begin{equation}
G^{\alpha}(x,\zeta)=\Gamma_{\beta\gamma}^{\alpha}\zeta^{\alpha}\zeta^{\gamma
}+\mathcal{B}_{\beta}^{\alpha}\zeta^{\beta}+\zeta^{\alpha}\,Z\,d(Z^{-1})/ds.
\label{11/-}%
\end{equation}
In these equations $\Gamma_{\beta\gamma}^{\alpha}$ are the Christoffel symbols
in Minkowski space-time,
\[
Z^{-1}=d\sigma/ds=(\eta_{a\beta}\zeta^{\alpha}\zeta^{\beta})^{1/2},\;
\mathcal{B}_{\beta}^{\alpha}=\eta^{\alpha\delta}\mathcal{B}_{\delta\beta}%
\]
and
\[
\mathcal{B}_{\delta\beta}=\partial f_{\beta}/\partial x^{\delta}-\partial
f_{\delta}/\partial x^{\beta}.
\]

In Finslerian space-time a number of connections compatible with 
(\ref{11/29}) can be defined \cite{Rund}. In particular, this equation can be
interpreted in the sense that in NIFRs space-time the absolute derivative of a
vector field $\xi^{a}(x)$ along the world line $x^{\alpha}=x^{\alpha}(s)$ is
given by
\begin{equation}
D\xi^{\alpha}/ds=d\xi^{\alpha}/ds+G_{\beta}^{\alpha}(x,dx/ds)\xi^{\beta},
\label{11/30}%
\end{equation}
where
\[
G_{\beta}^{\alpha}(x,dx/ds)=\mathcal{B}_{\beta}^{\alpha}+\Gamma_{\beta\gamma
}^{\alpha}\, dx^{\gamma}/ds+Z\,dZ^{-1}/ds.
  \]

Equations (\ref{11/30}) define a connection of the Laugwitz type \cite{Rund}
in the space-time of a NIFR, which is nonlinear with respect to $dx^{\alpha}$
. The change in the vector $\xi^{\alpha}$ due to an infinitesimal parallel
transport is
\begin{equation}
\label{11-1}d\xi^{\alpha} = - G^{\alpha}_{\beta}(x,dx) \xi^{\beta} ,
\end{equation}

Consider a free motion of a particle of mass $m$ in a NIFR. Since
$ds=d\sigma-f_{\alpha}dx^{\alpha}$,  equations of the motion are
\begin{equation}
Du^{\alpha}/ds=\mathcal{B}_{\beta}^{\alpha}u^{\beta}. \label{11/31}%
\end{equation}

The equations of the motion (\ref{11/29}) on a non-relativistic disk rotating in the
plane $xy$ about the axis $z $ with an angular velocity $\Omega$ are take the
form 
\begin{equation}
d\vec{\zeta}/dt+\vec{\Omega}\times\vec{r}=0, \label{11/32},%
\end{equation}
where $\vec{r}=\{x,y,z\}$ and the coordinates origin coincides with the disk
center. The absolute derivative (\ref{11/30}) of a vector $\vec{\xi}$ is given
by
\begin{equation}
D\vec{\xi}/dt=d\vec{\xi}/dt-\vec{\Omega}\times\vec{\xi}. \label{12/33}%
\end{equation}
and  equations of the motion (\ref{11/31}) of the considered particle in the
NIFR are
\begin{equation}
D\vec{v}/dt=-\vec{\Omega}\times\vec{v}, \label{12/34}%
\end{equation}
where $\vec{v}=\{\dot{x}\, \dot{y},\, \dot{z}\}$.

Next, for the 4-velocity $u^{\alpha}$ we have
\begin{equation}
u^{\alpha}=\overline{u}^{\alpha}+\zeta^{\alpha}, \label{12/35}%
\end{equation}
where $\overline{u}^{\alpha}$ is the spatial velocity of the particle in the
NIFR. In the non-relativistic limit  (\ref{12/35}) can be written in the
form
\begin{equation}
\vec{v}=\overline{\vec{v}}+\vec{\zeta}, \label{12/36}%
\end{equation}
where $\overline{\vec{v}}$ is the ``relative velocity`` of the particle and
$\vec{\zeta}$ is the velocities field of the  points of the disk in the laboratory
frame. Substituting (\ref{12/36}) for (\ref{12/34}), we find that
\begin{equation}
D\overline{\vec{v}}/dt=-D\overline{\vec{\zeta}}/dt-\vec{\Omega}\times\vec{v}
-\vec{\Omega}\times\vec{\zeta}. \label{12/37}%
\end{equation}

The value $D\overline{\vec{v}}/dt$ is an acceleration of the considered
particle as measured by an observer in the NIFR. The velocities field $\vec{
\zeta}$ of the disk points is given by
\begin{equation}
\vec{\zeta} = \vec{\Omega} \times\vec{r}.
\end{equation}

Hence, along the particle path we have $d\vec{\zeta} /dt = \vec{\Omega}
\times\vec{v}$ and
\begin{equation}
\label{12/38}D \vec{\zeta} /dt = d\vec{\zeta}/dt - \vec{\Omega} \times
\vec{\zeta} = \vec{ \Omega} \times\overline{\vec{v}} .
\end{equation}

Therefore, finally, we find from (\ref{12/34})
\begin{equation}
mD\overline{\vec{v}}/dt=-2m(\vec{\Omega}\times\overline{\vec{v}})-m\vec{
\Omega}\times(\vec{\Omega}\times\vec{r}). \label{12/39}%
\end{equation}
We have arrived at non-relativistic equations of motion of a particle in
rotating frames \cite{Syng}. The right-hand side of  (\ref{12/39}) is the
expression for the Coriolis and centrifugal forces.

Thus, in non-relativistic limit the Finslerian space-time in NIFRs manifests
itself in the structure of the derivative of vectors with respect to time $t$
, see also \cite{Verozub81b}.

It is interesting to note that  (\ref{12/33}) is considered sometimes in
classical dynamics nominally \cite{Syng} just to obtain the expression for
inertial forces in NIFRs.

\section{Relativity of Inertia}

A clock, which is in a NIFR at rest, is unaffected by acceleration in
space-time of this frame. A difference in the rate of an ideal clock in IFRs
and NIFRs is a real consequence of a difference between the space-time metric
in the IFR and NIFR. It is given by the factor $Z=ds/d\sigma$. For a rotating
disk of the radius $R$ the factor  $Z=1-\omega^{2}R^{2}/2c^{2}$ which gives rise to the
observed redshift in the well-known Pound-Rebka-Snider experiments.

Consider another experimentally verifiable consequence of the above theory.
Let $p^{\alpha}=mc\ dx^{\alpha}/d\sigma$ be 4-momentum of a particle in an
IFR. From the point of view of an observer in a NIFR
\begin{equation}
p^{\alpha}=\overline{p}^{\alpha}+c^{-1}E\,\zeta^{\alpha}.
\end{equation}
In this equations the spatial projection $\overline{p}^{\alpha}$ should be
identified with the momentum and the quantity $E$ with the energy of the particle.

It is obvious that $E=\overset{\ast}{\zeta}_{\alpha}p^{\alpha}$. Therefore,
the energy of the particle in the NIFR is
\begin{equation}
E=m\,Z\,c^{2}\overset{\ast}{\zeta}_{\alpha}u^{\alpha},
\end{equation}
where $Z=ds/d\sigma=F(x,dx/d\sigma)$. For the particle at rest in the NIFR
$u^{\alpha}=\zeta^{\alpha}$ and we obtain
\begin{equation}
E=m\,Z\,c^{2}.%
\end{equation}
Thus, the inertial mass $m_{1}$ of the particle in the NIFR is given by
\begin{equation}
m_{1}=Z\,m.
\end{equation}
The quantity $m$ coincides with the proportionality factor between the
momentum and the velocity of a nonrelativistic particle in the NIFR.

Since $Z$ is the function of $x^{\alpha}$, the inertial mass $m_{n}$ in the
NIFR is not a constant. For example, on a rotating disk we have
\begin{equation}
m_{n}=m\,/(1-\Omega^{2}r^{2}/2c^{2}),
\end{equation}
where $\Omega$ is the  angular velocity and $r$ is the distance of the
body from the disk centre.  The difference between the inertial mass $m^{eq}$ of
a body on the Earth equator and the mass $m^{pol}$ of the same body on the
pole is given by
\begin{equation}
(m^{eq}-m^{pol})/m^{pol}=1.2\cdot10^{-12}.%
\end{equation}

The dependence of the inertial mass of particles on the Earth longitude can be
observed by the M$\ddot{\text{o}}$ssbauer effect. Indeed, the change
$\Delta\lambda$ in a wave length $\lambda$ at the Compton scattering on
particles of the masses $m$ is proportional to $m$. If this value is measured
for gamma-quanta with the help of the M$\ddot{\text{o}}$ssbauer effect at a
fixed scattering angle, then after transporting the measuring device from the
longitude $\varphi_{1}$ to the longitude $\varphi_{2}$ we obtain
\begin{equation}
\frac{\Delta\lambda_{\varphi_{1}}^{-1}-\Delta\lambda_{\varphi_{2}}^{-1}}{
\Delta\lambda_{\varphi_{1}}^{-1}}=\Theta\ [\cos^{2}\varphi_{1}-\cos^{2}%
\varphi_{2}],
\end{equation}
where $\Theta=1.2\cdot10^{-12}$.

\section{Gravitation in Inertial and Proper Reference Frames}

To implement the the bimetric description of gravity based on the relativity of space-time to the employed reference frame , suppose \cite{Thirring}
that in Minkowski space-time gravitation can be described as a tensor field $
\psi_{\alpha\beta}(x)$ of spin 2, for which the Lagrangian, describing the
motion of a test particle of  mass $m$ is of the form 
\begin{equation}
L=-mc[g_{\alpha\beta}(\psi)\;\dot{x}^{\alpha}\;\dot{x}^{\beta}]^{1/2},
\label{LagrangianThirr}
\end{equation}
where $\dot{x}^{\alpha}=dx^{\alpha}/dt$ and $g_{\alpha\beta}$ is the
symmetric tensor whose components are functions of $\psi_{\alpha\beta}$.

Consider a frame of reference, the reference body of which is formed by
identical point masses $m$ moving in an IFR under the effect of the field $
\psi_{\alpha\beta}(x)$. It is a proper frame of reference of the given
field. An observer, located in the PFR at rest, moves along a geodesic line
of his space-time.

According to  (\ref{dsMain}) the square of the line element in a PFR
 of the field $\psi_{\alpha\beta}$ is given by 
\begin{equation}
ds^{2}=g_{\alpha\beta}(\psi)\;dx^{\alpha}\;dx^{\beta}.
\end{equation}

Thus, according to what has been outlined in Sec. 4, space-time in
PFRs is Riemannian $V_{4}$ with the curvature other than zero. Viewed by an
observer located in the IFR, the motion of the particles, forming the
reference body of the PFR, is affected by the force field $
\psi_{\alpha\beta} $. Let $x^{i}(t,\chi)$ be a set of paths of the motion of 
the  depending on the parameter $\chi$. Then, for the observer located in the IFR the
relative motion of a pair of particles from the set is described in
non-relativistic limit by the differential equations \cite{MTU} 
\begin{equation}
\frac{\partial^{2}n^{i}}{\partial t^{2}}+\frac{\partial^{2}U}{\partial
x^{i}\partial x^{k}}n^{k}=0,  \label{DevIFR}
\end{equation}
where $n^{k}=\partial x^{k}/\partial{\chi}$ and $U$ is the gravitational
potential.

However, the observer in a PFR of this field will not feel the existence of
the field since he moves in the space-time of the PFR along a geodesic line.
The presence of the field $\psi_{\alpha\beta}$ will be displayed for him
differently --- as space-time curvature which manifests itself as a deviation
of the world lines of nearby points of the reference body.

For a quantitative description of this fact it is natural for him to use the
Riemannian normal coordinates.\footnote{This and the above consideration does not depend on the used coordinate system, it can be performed by a covariant method.}
 In these coordinates spatial components of
the deviation equations of geodesic lines are 
\begin{equation}
\frac{\partial n^{i}}{\partial t^{2}}+R_{0k0}^{i}n^{k}=0,
\end{equation}
where $R_{0k0}^{i}$ are the components of the Riemann tensor. If the metric
tensor does not depend on time $t$, then 
\begin{equation}
R_{0k0}^{i}=-g^{ri}\frac{\partial^{2}g_{00}}{\partial x^{r}x^{k}}.
\end{equation}
In the Newtonian limit these equations coincide with (\ref{DevIFR}).

Thus, in two  frames of reference being used we have two different descriptions of particles 
motion --- as moving under the action of a force field in Mankowski space-time, and as moving 
along the geodesic line in a Riemann space-time
with the curvature other than zero.

Of course,  (\ref{dsMain}) refers to any classical field. For instance,
space-time in PFRs of an electromagnetic field is Finslerian. However, since 
$ds$, in this case, depends on the mass and charge  of the particles
forming the reference body, this fact is not of great significance.

If we start from the Lagrangian (\ref{LagrangianThirr}) for the motion of test particles, Einstein's
equations cannot be considered as equations for finding functions $g_{\alpha
\beta }(x)$. The reason is that the gravitational equations for $g_{\alpha
\beta }(\psi)$ should be some  differential equations which are invariant under a certain group
of gauge transformations $\psi _{\alpha \beta }(x)\rightarrow \overline{\psi 
}_{\alpha \beta }(x)$ which are a consequence of the existence of ''extra``
components of the tensor $\psi _{\alpha \beta }$. These transformations induce 
 some transformations  $g_{\alpha \beta }(x)\rightarrow \overline{g}_{\alpha \beta}(x)$.  
Equations of motion of particles  resulting from the Lagrangian (\ref{LagrangianThirr}) 
certainly should be invariant under such transformations
of the tensor $g_{\alpha \beta }(x)$. 
But the   equations of motion  are, at the same times, differential equations of
geodesic lines in  Riemannian's space-time  of a proper frame of reference. 
Consequently, if  equations of gravitation do not contain functions $\psi _{\alpha \beta }$
explicitly, i.e. are differential equations for $g_{\alpha \beta }(x)$,
they should be invariant under such transformations of  metric tensor of space-time in PRFs,  
 and  it takes place in any  coordinate system at that.

\section{Spherically-Symmetric Gravitational Field}

Let us find a spherically-symmetric vacuum solution $g_{\alpha\beta}(x)$ of bimetric
equations (\ref{myEqsWithMatter}) in Minkowski space-time $E_{4}$. Such a
solution, considered as a tensor field in $E_{4}$, describes, in particular, the
gravity of a point mass $M$ for a remote observer in an IFR for which
space-time is supposed to be pseudo-Euclidean. In this case, if the 
Lagrangian of a particle is invariant under the mapping $t\rightarrow
-t$, in the spherical coordinates of the Minkowski space-time,  it is of the
form
\begin{equation}
L=-m\,c\,[A\dot{r}^{2}+B(\dot{\theta}^{2}+\sin^{2}\theta\ \dot{\varphi}%
^{2})-c^{2}C]^{1/2} \label{LagrTestParticls},%
\end{equation}
where $A$, $B$ and $C$ are the functions of the radial coordinate $r$.

The associated line element of space-time  in PFRs is
\begin{equation}
ds^{2}  =A\,dr^{2}+B[\,d\theta^{2}+\sin^{2}\theta\,\,\varphi^{2}]
  -C\,{dx^{0}}^{2}, \label{ds}.%
\end{equation}

Some additional conditions can be imposed on the tensor $g_{\alpha\beta}$
because of the gauge (geodesic) invariance. In particular, under the
conditions $ Q_{\alpha}=0$
Eqs. (\ref{MyVacuumEqs}) are reduced to Einstein's vacuum equations
$R_{\alpha\beta}=0$.. Thus, the functions $A$, $B$ and $C$ can be found as a
solution of the system of the differential equations :
\begin{equation}
R_{\alpha\beta}=0 \label{Einstein Eqs}%
\end{equation}
and
\begin{equation}
Q_{\alpha}=0, \label{AditionalCondition}%
\end{equation}
which satisfy the conditions:
\begin{equation}
\lim\limits_{r\rightarrow\infty}A=1,\;\lim\limits_{r\rightarrow\infty}%
(B/r^{2})=1,\;\lim\limits_{r\rightarrow\infty}C=1. \label{limitConditions}%
\end{equation}
It allows to use all mathematical methods elaborated for general relativity.

Conditions (\ref{AditionalCondition}) yield one equation:
\begin{equation}
B^{2}AC=r^{4}.
\end{equation}
It allows to exclude the function $A$ from the (\ref{Einstein Eqs}). Then the
equations $R_{11}=0$ and $R_{00}=0$ are:%
\begin{equation}
-2BC^{\prime}+2rB^{\prime}C^{\prime}+rBC^{\prime\prime}=0, \label{EinstEq1}%
\end{equation}
\begin{equation}
\label{EinstEq2}-4BCB^{\prime}+rCB^{\prime2}-2BC^{\prime}+2rBB^{\prime
}C^{\prime}
+2rBCB^{\prime\prime}+rB^{2}C^{\prime\prime} = 0.
\end{equation}
Because of equality ( \ref{EinstEq1}) the sum of tree terms in (\ref{EinstEq2}%
) is equal to zero, and we obtain a differential equation for the function
$B$:
\begin{equation}
2rBB^{\prime\prime}+rB^{\prime2}-4BB^{\prime}=0.
\end{equation}
A general solution of this equation can be written as $B=a(r^{3}%
+\mathcal{K}^{3})^{2/3}$ where $a$ and $\mathcal{K}$ are some constants. The
constant $a=1$ which can be seen from condition (\ref{limitConditions}).

The function $C$ can be found from differential equation (\ref{EinstEq1}) in
the form
\begin{equation}
C^{\prime\prime}+2\frac{rB^{\prime}-B}{rB}C^{\prime}=0,
\end{equation}
where $(r B^{\prime}-B)/r B=(r^{3}-\mathcal{K}^{3})/(r^{4}+r \mathcal{K}^{3}%
)$. A general solution of this equation is $C=b-\mathcal{Q}/B^{1/2}$ where $b$
and $\mathcal{Q}$ are constant. It follows from (\ref{limitConditions}) that
$b=1$.

Thus, in the spherical coordinate system the functions $A$, $B$ and $C$ are
given by:
\begin{equation}
A=\frac{r^{4}}{f^{4}(1-{\mathcal{Q}}/f)},\;B=f^{2},\;C=1-\frac{\mathcal{Q}}%
{f}. \label{ABCequal},%
\end{equation}
where
\begin{equation}
f=(r^{3}+{\mathcal{K}}^{3})^{1/3}.%
\end{equation}

The nonzero components of the tensor $B_{\alpha\beta}^{\gamma}$ are given by%
\begin{align}
B_{rr}^{r} =\frac{1}{2}\frac{A^{\prime}}{A};\quad B_{\theta\theta}^{r}%
=\frac{1}{2}\frac{2rA-B^{\prime}}{A};
\quad B_{\phi\phi}^{r} =-\frac{1}{2}\frac{2rA-B^{\prime}}{A}\sin^{2}%
\theta; \nonumber\\
B_{tt}^{r} =\frac{1}{2}\frac{C^{\prime}}{A};\quad B_{tr}^{t}=\frac{1}{2}%
\frac{C^{\prime}}{C};
\quad B_{\theta r}^{\theta} =\frac{1}{2}\frac{2B-B^{\prime}r}{rB};\quad
B_{\phi r}^{\phi}=\frac{2B-B^{\prime}r}{rB}.\nonumber
  \end{align}

In non-relativistic limit the radial component of the equations of the motion
of a test particle (\ref{EqMotionOfTestPart}) takes the form $\ddot{x}%
^{r}=-c^{2}\,\Gamma_{00}^{r},$ where $\Gamma_{00}^{r}=C^{\prime}%
/2A=r^{4}C^{\prime}/f^{4}C.$ Therefore, to obtain the Newton gravity law it
should be supposed that at large $r$ the function $f(r)$ coincides with $r$,
and ${\mathcal{Q}}=r_{g}=2G\,M/c^{2}$ is the classical Schwarzschild radius.

At the given constant ${\mathcal{Q}}$ allowable solutions (\ref{ABCequal})
are obtained by
changing the arbitrary constant $\mathcal{K}$. In particular, if we set
$\mathcal{K}=0$, then the line element (\ref{ds})
of the space-time in PFRs coincides
with the Droste-Weyl solution of Einstein's equations \cite{Abrams1} (it is
commonly named Schwarzschild's solution) which has an event horizon at
$r=r_{g}$:%
\begin{equation}
ds^{2}=-\;\frac{dr^{2}}{(1-r_{g}/r)}-r^{2}[d\theta^{2}+\sin^{2}\theta
\;d\varphi^{2}]+(1-r_{g}/r)\;{dx^{0}}^{2}. \label{Drost}%
\end{equation}
If we set $\mathcal{K}={\mathcal{Q}}$, the line element  coincides with the
original Schwarzschild solution \cite{Schwarzschild}%
\begin{equation}
ds^{2}=-\;\frac{f^{\prime2}dr^{2}}{(1-r_{g}/f)}-f^{2}[d\theta^{2}+\sin
^{2}\theta\;d\varphi^{2}]+(1-r_{g}/f)\;{dx^{0}}^{2},
\label{MyMetricFormOfSpaceTime}%
\end{equation}
where $f=($ $r_{g}^{3}$ $+$ $r^{3})^{1/3}$. This solution has no  event
horizon and no singularity in the centre.
Really, for motion of a particle in the  plane $\theta=\pi/2$ 
 laws of conservation of energy $E$ and  momentum $J$ take place:
\begin{equation}
\dot{r}\frac{\partial L}{\partial\dot{r}}-\dot{\varphi}\frac{\partial
L}{\partial\dot{\varphi}}-L=E,  \, \, \, 
\frac{\partial L}{\partial\dot{\varphi}}=J.
\label{EnergyConservation}%
\end{equation}
It follows from this fact that the  equations of the motion are of the form
\begin{equation}
{\dot{r}}^{2}=(c^{2}C/A)[1-(C/\overline{E}^{2})(1+r_{g}^{2}\overline{J}%
^{2}/B)], \label{EqsMotionTestPart1},%
\end{equation}%
\begin{equation}
\dot{\varphi}=c\;C\overline{J}/r_{g}B\overline{E} \label{EqsMotionTestPart2},%
\end{equation}
where $(t,r,\varphi)$ are the spherical coordinates, $\dot{r}=dr/dt$,
$\dot{\varphi}=d\varphi/dt$ , $\overline{E}=E/mc^{2}$, $\overline{J}%
=J/r_{g}mc$. 

\begin{figure}[h]
\includegraphics[width=9cm,height=6cm]{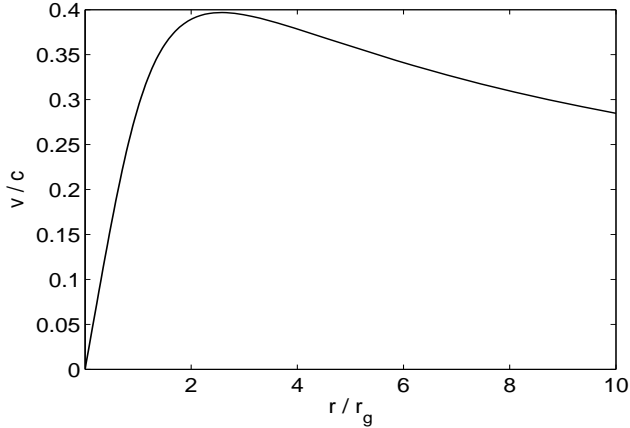} 
\caption{The velocity of a
freely falling particle (in units of $c$) as the function of the distance $r/r_{g}$
from the centre.}%
\label{VelocityFreePart}%
\end{figure}

The radial velocity of freely falling particle is given by the equation%
\begin{equation}
v=c\left[  \frac{C}{A}(1-C)\right]  ^{1/2}%
\end{equation}
Fig.~ \ref{VelocityFreePart} shows the plot of the velocity as a function of
the distance $r/r_{g}$ from the centre of a point attracting mass.
It follows from this figure that this function is defined at all distances from the centre,
and tends to zero at $r \rightarrow 0$.

Line elements  (\ref{Drost}) and (\ref{MyMetricFormOfSpaceTime}) in $V_{4}$
are not equivalent
because they were obtained in the same coordinate system and correspond to the
different values of the constant $\mathcal{K}$ in solution (\ref{ABCequal}).
 Both of the line elements refer
to the spherical coordinate system, defined in $E_{4}$, where $t$, $r$,
$\varphi$ and $\theta$ are magnitudes measured by measurement instruments.
Suppose, as it is usually believed,  (\ref{MyMetricFormOfSpaceTime}) can
be transformed to the form (\ref{Drost}) by an appropriate local coordinate
transformation $r\rightarrow r^{\prime}$ and $t\rightarrow t^{\prime}$.
\footnote{Problems of correctness of such transformation have been considered
in \cite{Abrams1}, and criticism with respect to the Droste-weyl solution in General
Relativity --- in \cite{Crothers}} and \cite{Liebscher}. But these new coordinates have not an
operational meaning, and after an appropriate transformation of the physical
3-space and time intervals, the same physical results will be obtained from
(\ref{Drost}) as from (\ref{MyMetricFormOfSpaceTime}). (Just as in classical
electrodynamics in arbitrary coordinates). In particular, since the line element (\ref{MyMetricFormOfSpaceTime}) at
$\mathcal{K}={\mathcal{Q}}$     has no the event horizon 
and the singularity in the centre in  spherical coordinates, 
it does not contain it in other coordinate systems.

It can be argued that the constant ${\mathcal{K}}$ must be equal 
to $\mathcal{Q} = r_{g}$.
Gauge-invariant tensor  $B_{\beta\gamma}^{\alpha}(x)$ can be considered as the field strength tensor of gravitational field.
There is a simple possibility to consider the object $B_{\beta\gamma}^{\alpha
}(x)$ as a function of a tensor field $\psi_{\alpha\beta}(x)$ which may be
interpreted as a potential $\psi_{\alpha\beta}(x)$ described gravity in $E_{4}$ as a field of spin two.
Namely, one can set%
\begin{equation}
B_{\beta\gamma}^{\alpha} =\nabla^{\alpha}\psi_{\beta\gamma} -(n+1)^{-1}\left(
\delta_{\beta}^{\alpha}\nabla^{\sigma}\psi_{\sigma\gamma}+\delta_{\gamma
}^{\alpha}\nabla^{\sigma}\psi_{\beta\sigma}\right)  . \nonumber\label{BbyPsi}%
\end{equation}
The identity $B_{\alpha\gamma}^{\gamma}=0$ is satisfied as it is to be
expected according to the definition of the tensor $B_{\alpha\gamma}^{\gamma}
$. Then, at the gauge condition $\nabla^{\sigma}\psi_{\sigma\gamma}=0$ Eqs.
(\ref{MyVacuumEqs}) are%
\begin{equation}
\label{EqsPsiVac}%
\square\psi_{\alpha\beta} = \nabla^{\sigma}\psi_{\alpha\gamma}\nabla
^{\gamma}\psi_{\sigma\beta}, \, \, \, \, \, 
\nabla^{\sigma}\psi_{\sigma\gamma} =0,
\end{equation}
where $\square$ is the covariant Dalamber operator in Minkowski space-time.

It is natural to suppose that in the presence of matter these equations take
the form\footnote{It should be noted that, when we introduce $\psi
_{\alpha\beta}$ in such a way, we cannot be sure a priori that the equations
for $\psi_{\alpha\beta}$ yield all physical solutions of the equations for
$B_{\beta\gamma}^{\alpha}$.}
\begin{align}
\label{EqPsiMatter}%
\begin{split}
\square\psi_{\alpha\beta} =\varkappa(T_{\alpha\beta}+t_{\alpha\beta}),\\
\nabla^{\sigma}\psi_{\sigma\gamma} =0,
\end{split}
\end{align}
where $\varkappa=8\pi G/c^{4}$,
\begin{equation}
t_{\alpha\beta}= \varkappa^{-1} B_{\alpha\mu}^{\nu}\;B_{\beta\nu}^{\mu} =
\varkappa^{-1}\nabla^{\sigma}\psi_{\alpha\gamma}\,\nabla^{\gamma}\psi
_{\sigma\beta}, \label{TikGravField}%
\end{equation}
and $T_{\alpha\beta}$ is  the energy-momentum tensor of matter.

Obviously, the equality
\begin{equation}
\nabla^{\beta}(T_{\alpha\beta}+t_{\alpha\beta})=0 \label{ConservLawPsi}%
\end{equation}
is valid. Therefore, the magnitude $t_{\alpha\beta}$ can be interpreted as the
energy-momentum tensor of the gravitational field.

Let us find the energy of the gravitational field of a point mass $M$ as an
integral in Minkowski space-time:
\begin{equation}
{\mathcal{E}}_{gf}=\int t_{00}dV, \label{energydef}%
\end{equation}
where $dV$ is a volume element. In Newtonian theory this integral is
divergent. In our case we have:
\begin{equation}
t_{00}=2\varkappa^{-1}B_{00}^{1}\;B_{01}^{0}=\frac{c^{4}}{8\pi G}%
\frac{\mathcal{Q}^{2}}{f^{4} \,\mathcal{K}} 
\label{t00}%
\end{equation}
and, therefore, in  spherical coordinates, we obtain%
\begin{equation}
{\mathcal{E}}_{gf}=\int t_{00}dV=\frac{1}{8}\frac{{\mathcal{Q}}%
^{2}c^{4}}{\pi G}\, \mathcal{J}, \label{energycalc}%
\end{equation}
where \cite{Byrd}%
\begin{equation}
\mathcal{J}=\int\frac{dV}{f^{4}}=\frac{4\pi}{3{\mathcal{K}}}B(1,1/3)
\end{equation}
and
\begin{equation}
B(z,w)=\int_{0}^{\infty}\frac{t^{z-1}}{(1+t)^{z+w}}dt \label{B-function}%
\end{equation}
is the B-function. Using the equality
\begin{equation}
B(z,w)=\frac{\Gamma(z)\Gamma(w)}{\Gamma(z+w)}, \label{BataGamma}%
\end{equation}
where $\Gamma$ is the $\Gamma$-function, we obtain $\mathcal{J}=4\pi/{\mathcal{K}%
}$, and, therefore,
\begin{equation}
{\mathcal{E}}_{gf}=\frac{{\mathcal{Q}}}{{\mathcal{K}}}\frac{{\mathcal{Q}}%
c^{4}}{2G}=\frac{{\mathcal{Q}}}{{\mathcal{K}}}M\,c^{2} \label{energyfinally}.%
\end{equation}

We arrive at the conclusion that at ${\mathcal{K}}\neq0$ the energy of a
point-mass is finite and, in particular, \ at $\mathcal{K}=\mathcal{Q}$ it is
caused entirely by its gravitational field:
\begin{equation}
\mathcal{E}_{gf}=M\,c^{2}.
\end{equation}

The spatial components of the vector $P_{\alpha}=t_{0\alpha}$ are equal to zero.

Due to these facts we assume in the present paper that ${\mathcal{K}%
}={\mathcal{Q}}=r_{g}$ and consider the functions
\begin{equation}
\label{ABCfinal}
\begin{split}
A  &  =\frac{r^{4}}{f^{4}(1-{\mathcal{Q}}/f)},\ C=1-r_{g}/f,\ B=f^{2};\\
f  &  =(r_{g}^{3}+r^{3})^{1/3}%
\end{split}
\end{equation}
in the Lagrangian (\ref{LagrTestParticls}) as a basis for  analysis of physical effects.

It must be noted that by using the above results we can find also the gravitational field inside a 
spherically-symmetric matter layer. In order to reach a coincidence of the equation of motion  of test 
particles in nonrelativistic limit with the Newtonian one, the constant $\mathcal{Q}$ in  (\ref{ABCequal})
 in this case must be equal to zero. Therefore, the spherically-symmetric matter layer does not create gravitational field inside itself. This result will be used in next section.

As an observer, located in a PFR does not feel influence of any forces, he
should explain the fact of the deviation of paths of close particles of the
PFR reference body as a manifestation of deviation of geodesic lines of these
point masses in space-time with the line element 
(\ref{MyMetricFormOfSpaceTime}). However, unlike the situation in General Relativity,
 he does not
observe any problems at the Schwarzschild radius and close to $r=0$ because
components of the curvature tensor have not any singularity. For example
$R_{010}^{1}=r_{g}f$ $^{-4}(r_{g}-f)$. It  tends to zero when $r\rightarrow
0$.  This fact is in accordance with the fact that in flat space-time the
gravitational force tends to zero when the distance from the central point
mass  tends to zero, see next section.

\section{ Cosmological test}

Physical consequences from the spherical-symmetric solution of equations (\ref{myEqsWithMatter}) differ 
very little  from those in general relativity at distances $r \gg r_{g}$
but they are different in principle  at  distances of the order of the Schwarzschild
radius or less than that. 
Some principal physical consequences 
were considered in papers \cite{Verozub96}, \cite{VerKochStability}, \cite{Verozub04}.
       Some of them have chances to be checked up.
The equations (\ref{myEqsWithMatter})  were successfully tested by the binary
pulsar $PSR1913+16$  \cite{verkoch00}.  Here
we consider another problem.

Within last 8 years numerous data were obtained testifying that the most distant galaxies
move away from us with acceleration  \cite{riess}.
This fact poses serious problems both for astrophysics and for fundamental
physics  \cite{Weinberg}. In the present paper it is shown that
available observed data are an inevitable consequence of properties of the
gravitational force as deduced from the geodesic-invariant bimetric gravitation equations.

According to  (\ref{EqMotionOfTestPart}),  the gravitational force $m\, \ddot{r}$ of a
point mass $M$ affecting a freely falling particle of mass $m$ in Minkowski
space-times is given by
\begin{equation}
F=-m\left[  c^{2}C^{\prime}/2A+(A^{\prime}/2A-2C^{\prime}/2C)\dot{r}%
^{2}\right],
 \label{gravaccel1},%
\end{equation}
where the functions $A$ and $C$ are given by Eqs. (\ref{ABCfinal} ) and $r$ is the distance
from the centre.

\begin{figure}[h]
\includegraphics[width=9cm,height=6cm]{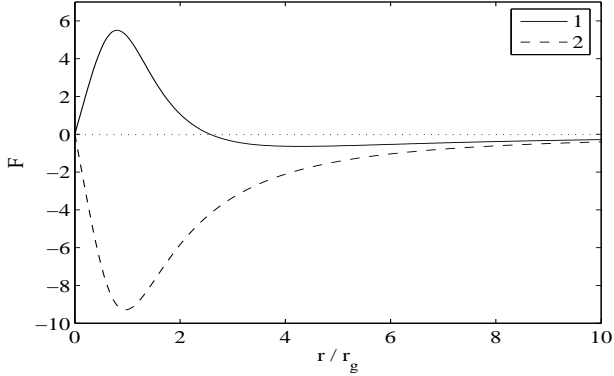}\caption{The gravitational force (arbitrary
units) affecting freely  particles (curve 1) and particles at rest (curve
2) near an attractive point mass.}%
\label{Force}
\end{figure}

For particles at rest ($\dot{r}=0$)
\begin{equation}
F=-\frac{GmM}{r^{2}}\left[  1-\frac{r_{\mathrm{g}}}{(r^{3}+r_{\mathrm{g}}%
^{3})^{1/3}}\right]  \label{ForceStat}.%
\end{equation}
Fig. \ref{Force} shows the force $F$ affecting particles at rest and the ones
, free falling from infinity on a point mass, as the function of the distance
$\overline{r}=r/r_{\mathrm{g}}$ from the centre.

It follows from this figure that the main peculiarity of the force acting on particles at rest in Minkowski space-time 
is that it tends to zero at $r \rightarrow 0$.
The gravitational force acting on freely moving
particles essentially differs from that acting on particles at rest. 
In this case the force changes its sign at $\sim2r_{g}$ and becomes
repulsive. Although we have so far never observed the motion of particles  at distances
of the order of $r_{g}$, we can verify this result for very remote objects in
the Universe at large cosmological redshifts, because it is well-known that
the radius of the observed region of the Universe is of the order of its
Schwarzschild radius.

A magnitude which is related to observations of the expanding Universe is a 
  velocity of distant star objects  moving off from the observer. 
To find the radial velocity of such objects  which is at the distance $R$ from the
observer, we can use spherically-symmetric  solution (\ref{ABCfinal}) of the vacuum
equations (\ref{MyVacuumEqs}) taking into account the effect of gravity of a spherically symmetric matter layer, as pointed out at the end of Sec. 8.
 In this case $M=(4/3)\pi\rho R^{3}$ is the
matter mass inside the sphere of the radius $R$, $\rho$ is the observed matter
density, and $r_{g}=(8/3)\pi c^{-2}G\rho R^{3}$ is Schwarzschild's radius of
the matter inside of the sphere. 

\begin{figure}[htb]
\includegraphics[width=9cm,height=6cm]{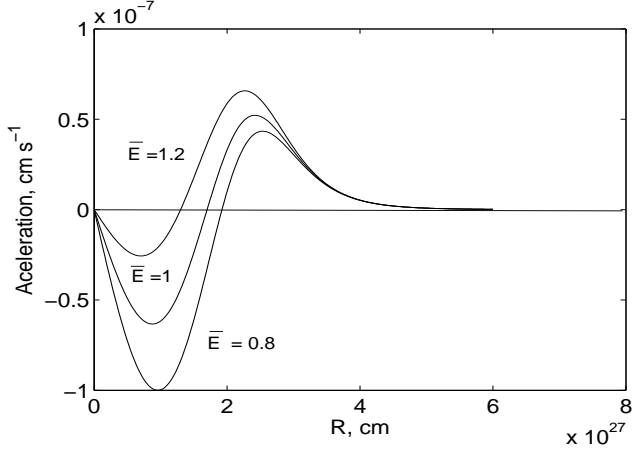}
\caption{The 
acceleration of particles on the surface of a homogeneous sphere of  radius $R$ for
three value of the parameter $\overline{E}$. The matter density is equal to
$10^{-29}{g}\,{cm}^{-3}$.}%
\label{fig:aceleration}%
\end{figure}

There is no necessity to demand the global
spherical symmetry of matter outside of the sphere. Indeed, suppose, the
matter density $\rho$ at the moment is of the order of the observed density of
the matter in the Universe ($10^{-29}\, - \, 10^{-30}\, {g}\,{cm}^{-3}$). Fig.
\ref{fig:aceleration} shows the radial acceleration $\ddot{R}$ of a particle
in the expanding Universe on the surface of the sphere of the radius $R$.

Two conclusions can be drawn from this figure.\newline\noindent1. At some
distance from the observer the relative acceleration changes its sign. If the
$R<2\cdot10^{27}{cm}$,  radial acceleration of particles is negative. If
$R>2\cdot10^{27}{cm}$, it magnitute is positive. Hence, for sufficiently
large distances the gravitational force affecting particles is repulsive and
gives rise to a relative acceleration of particles.

\noindent
2. The gravitational force, affecting the particles, tends to zero when $R$
tends to infinity. The same fact takes place as regards the force acting on
particles in the case of a static matter . The reason of the fact is that at a
sufficiently large distance $R$ from the observer the Schwarzschild radius of
the matter inside the sphere of  radius $R$ becomes of the same order as its
radius. Approximately at $R \sim2\, r_{g}$ the gravitational force begins to
decrease. The ratio $R/r_{g}$ tends to zero when $R$ tends to infinity, and
under these circumstances the gravitational force in the theory under
consideration tends to zero. Therefore, the gravitational influence of very
distant matter is negligible small.

Thus, according to (\ref{EqsMotionTestPart1}), the radial velocity of a star
object at the distance $R$ from the observer takes the form
\begin{equation}
v=c\frac{C\,f^{2}}{R^{2}}\sqrt{1-\frac{C}{\overline{E}^{2}}},
\label{StarVelocity}%
\end{equation}
where  $C$ and $f$    are  functions of distance $R$ of the object from an
observer, $v=dR/dt$, $\overline{E}$ is the total energy of a particle divided
by $mc^{2}$.

Proceeding from this equation we will find Hubble's diagram following mainly the method 
being used.in \cite{Zeldovich2}. 
Let $\nu_{0}$ be
a local frequency in the proper reference frame of a moving source at the
distance $R$ from the observer, $\nu_{l}$ be this frequency in a local
inertial frame, and $\nu$ be the frequency as measured by the observer in the
 centre of the sphere. The redshift $z=(\nu_{0}-\nu)/ \nu$ is caused by both
Doppler-effect and gravitational field. The Doppler-effect is a consequence of
a difference between the local frequency of the source in inertial and
comoving reference frame, and it is given by \cite{Landau}%
\begin{equation}
\label{DopplerRedshift}\nu_{l}=\nu_{0}\, [1-\sqrt{(1-v/c)(1+v/c)}].
\end{equation}

The gravitational redshift is caused by the matter inside the sphere of the
radius $R$. It is a consequence of the energy conservation for photon. The
energy integral (\ref{EnergyConservation}) for the motion of free particles
together with the Lagrangian (\ref{LagrTestParticls}) yield at $\dot{v}=0$ and
$\dot{\varphi}=0$ the rest energy of a particle in gravitational field:
\begin{equation}
E=m\,c^{2}\sqrt{C}.
\end{equation}
Therefore, the difference in two local level energy $E_{1}$ and $E_{2}$ of an
atom in the field is $\Delta E=(E_{2}-E_{1})\sqrt{C}$, so that the local
frequency $\nu_{0}$ at the distance $R$ from an observer are related with the
observed frequency $\nu$ by equality
\begin{equation}
\nu=\nu_{l}\sqrt{C}, \label{redsiftGrav},%
\end{equation}
where we take into account that for the observer location $\sqrt{C}=1$. It
follows from (\ref{redsiftGrav}) and (\ref{DopplerRedshift}) that the
relationship between the frequency $\nu$ as measured by the observer and the
proper frequency $\nu_{0}$ of the moving source in the gravitational field
takes the form
\begin{equation}
\frac{\nu}{\nu_{0}}=\sqrt{C\,\frac{1-v/c}{1+v/c}} \label{TotalRedshift},%
\end{equation}
which in the newtonian limit is of the form
\begin{equation}
\nu=\left(  1-\frac{v}{c}-\frac{4}{3}\frac{\pi G\rho R^{2}}{c^{2}}\right)
\nu_{0}.
\end{equation}

Equation (\ref{TotalRedshift}) yields the quantity $z=\nu_{0}/\nu-1$ as a
function of $R$. By solving this equation numerically we obtain the dependence
$R=R(z)$ of the measured distance $R$ as a function of the redshift. Therefore
the distance modulus to a star object is given by
\begin{equation}
\label{muofz}\mu=5 log_{10}[R(z)\, (z+1)] -5.
\end{equation}
where $R(1+z)$ is a bolometric distance (in $pc$) to the object.

If we demand that  (\ref{StarVelocity}) has to give a correct radial
velocity of distant star objects in the expansive Universe, it has to lead to
the Hubble law at small distances R. At this condition the Schwarzschild
radius $r_{g}=(8/3) \pi G \rho R^{3}$ of the matter inside the sphere is very
small compared with $R$. For this reason $f\approx r$, and $C=1-r_{g}/r$.
Therefore, at $\overline{E}=1$, we obtain from (\ref{StarVelocity}) that
\begin{equation}
v=H R, \label{Hubble}%
\end{equation}
where
\begin{equation}
\label{HubbleConstFormula}H=\sqrt{(8/3) \pi G \rho}.
\end{equation}
\begin{figure}[hbp]
 \includegraphics[width=9cm,height=6cm]{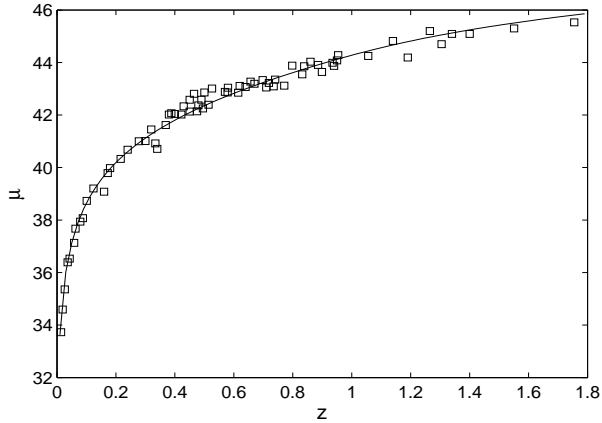}
\caption{The
distance modulus $\mu$ vs. redshift $z$ for the density $\rho=4.5
\cdot10^{-30} g\, cm^{-3}$. Small squares denote the observation data
according to Riess et al.}%
\label{fig:HubbleFig}%
\end{figure}
If $\overline{E}\neq1$ equation (\ref{StarVelocity}) does not lead to the
Hubble law since $v$ does not tend to $0$ when $R\rightarrow0$. For this
reason we set $\overline{E}= 1$ and look for the value of the density at which
a good accordance with observation data can be obtained.
 Fig.~ (\ref{fig:HubbleFig}) show the Hubble diagram obtained by 
(\ref{muofz}) compared with the Riess et al data \cite{riess} .
It follows from this figure that the model under consideration
does not contradict observation data. With the value of the density $\rho=4.5
\cdot10^{-30} g\, cm^{-3}$ we obtain from (\ref{HubbleConstFormula}) that
\begin{equation}
H=1.59\cdot10^{-18} c^{-1}=49\, (km/s)/ \, Mpc.
\end{equation}

\begin{figure}[tbp]
\includegraphics[width=9cm,height=6cm]{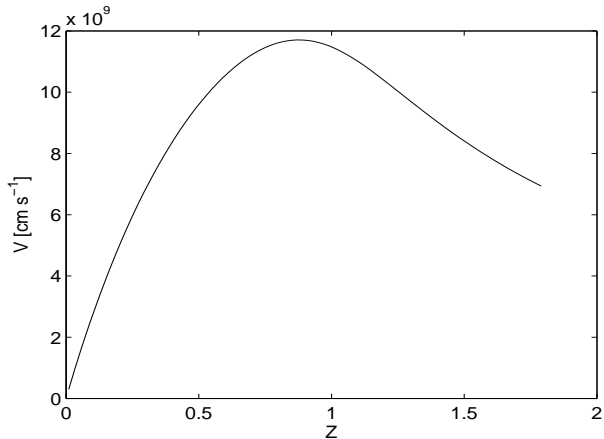}\caption{The radial velocity
vs. redshift $z$ for the density $\rho=4.5 \cdot10^{-30} g\, cm^{-3}$}%
\label{fig:VvsZ}%
\end{figure}

Fig. \ref{fig:VvsZ} shows the dependence of the radial velocity $v$ on the
redshift. It follows from this figure that at $z> 1$ the Universe expands with
an acceleration. At $R\rightarrow\infty$ the velocity and acceleration tend to zero.

\section{Conclusion}
 The necessity of a discussion of  the physical description of gravity through geodesic-invariant equations is quite obvious. However, the satisfactory implementation of such a programme is a difficult task. 
   The equations of gravitation considered in the given paper do not contradict observation data, and their some predictions seem rather attractive.     However it is still unclear  whether they are  only possible equations. 
    Besides, it appears that the correct equations of this type should be the bimetric ones. 
     However physical interpretation of such bimetricity involves    deep problems of the space-time theory which presently have no  a satisfactory understanding and a complete solution.
 

\end{document}